# A new 130nm F.E readout chip for Si microstrip detectors

Aurore Savoy-Navarro[1], Jean F. Genat[1], Th. Hung Pham[1], Rachid Sefri[1], Albert Comerma[2], Angel Dieguez[2] *

1 – LPNHE – Université Pierre et Marie Curie/IN2P3-CNRS
4, Place Jussieu, 75252 Paris-Cedex05 – France

2 – Universitat de Barcelona – Department d'Estructura i Constituents de la Materia,
Avenida Diagonal, Barcelona – Spain

In the context of the Silicon tracking for a Linear Collider (SiLC) R&D collaboration, a highly compact mixed-signal chip has been designed in 130nm CMOS technology intended to read Silicon strip detectors for the experiments at the future International Linear Collider. The chip includes eighty eight channels of a full analog signal processing chain and analog to digital conversion with the corresponding digital controls and readout channels. The chip is 5x10mm$^2$ where the analog implementation represents 4/5 of the total Silicon area.

## 1 Introduction

Within the framework of the SiLC R&D collaboration [1] an ASIC in deep submicron electronics (currently 130nm CMOS UMC technology) aiming to fully process the signals from micro-strip devices at the ILC is under development. The results obtained with a first prototype that includes the analog processing up and including digitization are briefly presented. The new version just back from foundry is a mix-mode chip with full analog and digital functionalities. The design and the main chip components are briefly described. This chip is now under extensive testing.

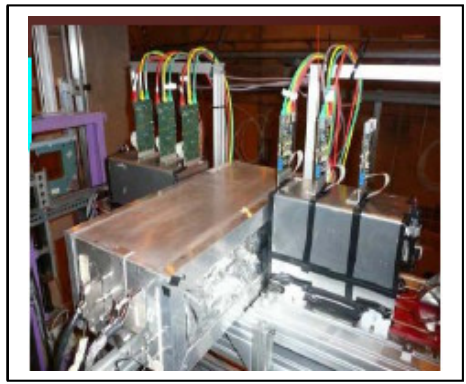

**Figure 1:** CERN Test beam with Si prototype equipped with SiTR_130-4 chips

## 2 Results with a preliminary prototype

A prototype with analog signal processing including the analog to digital conversion was tested at the Lab test bench with performances in agreement with expectations, i.e.: linearity of the preamplifier plus shaper up to 1% on the dynamic range up to 20 MIPs with 29 mV/MIP of gain, a peaking time between 0.8 to 2.5μs and a power dissipation of 0.6 mW per channel. Three Silicon micro-strips modules within a Faraday cage, read out by this preliminary chip version, were tested at a 120 GeV pion beam at CERN-SPS. They were located in between the two parts of the EUDET pixel telescope (Fig 1). The signal pulse height was reconstructed with a 16 cell analog sampling. A signal to noise ratio of 23 was achieved [2, 3] (Fig2).

* Acknowledgment to the EUDET E.U. project and to the SiLC R&D collaboration [1]

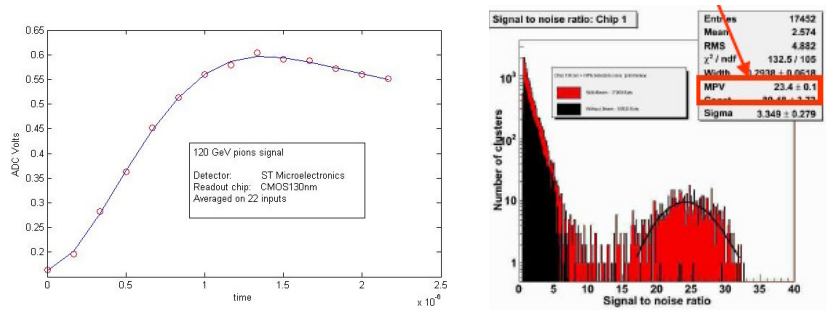

**Figure 2:** CERN Test beam results with Silicon strip prototype equipped with SiTR_130-4 chips: reconstructed pulse-height (left) and Signal to Noise ratio for a single chip (right).

## 3   Design and main features of the new SiTR_130-88 chip

The 88-channel prototype is a continuation of the 4-channel. Each channel includes a preamplifier-shaper, an 8x8-deep analog pipeline which samples the shaper output, a channel trigger decision on the sum of three adjacent channels (sparsifier) that freezes the pipeline and re-addresses the input to another pipeline. At the same time, the time information and channel index are collected. After data recording, pipeline outputs are converted in parallel using a single-ramp Wilkinson A/D converter. The converted data is associated with time and channel information before being serialized and sent to the output.  The digital implementation provides the chip with a high degree of flexibility and fault tolerance. The mix mode simulation is quite challenging with this complex circuit.

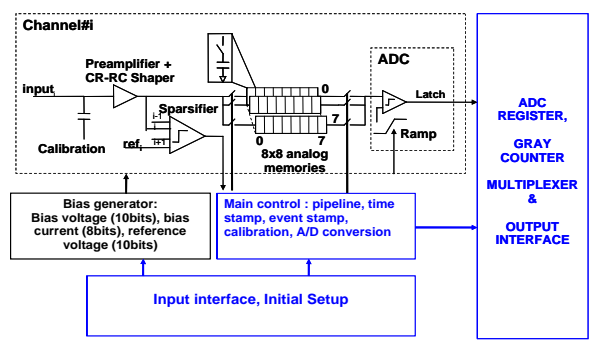

**Figure 3:** New 88-channel mix-mode chip, SiTR_130-88, flow diagram

The digital has been designed to match the analog. It has tree main blocs; control and configuration interface, acquisition control and readout bloc (Fig. 4). Between the three blocs there is a clear hierarchy, on the top there is the acquisition control which gives some control signals to the rest of the blocs permitting some operations. There is also some independence in operations between them since the control interface can operate at any time or any state and the readout block can be read at any time whenever it has relevant information.
The acquisition control bloc is constructed around a finite state machine (Fig. 4), with just four states: IDLE, the device is waiting for starting acquisition. In this state, the device can be powered off. START_PIPE, when the device comes from IDLE to acquisition, the first action is a specific sequence to initialize the pipelines. WRITE, once the initialization is achieved the relevant events are written in the analog pipelines. In this state some logic is triggered when a sparsifier response is detected. At the end, the READ state where the conversion to digital of the relevant data is performed, then the loop is repeated again until all the data stored in the analog pipelines has been converted and read. The chip was sent to foundry end of June 2008, received in October and under test since end of 2008. The chip covers 5x10mm$^2$ area with the 4/5 of this area taken by the analog part.

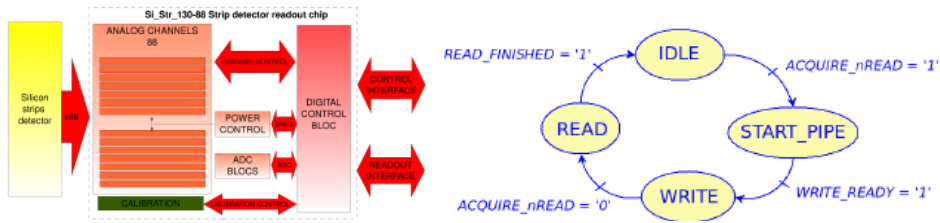

**Figure 4:** System Overview (left); Acquisition control Final State Machine (right)

## 4  What's next

The next goals are to design basic blocks of 256 channels each, and to gather these basic chips into a single device reading 1024 channels. The thinning of the chip is also under investigation and a version in 90nm is in the plan of work. Furthermore a version applied to the CLIC case (very fast cycling machine) is going to be developed. The development of the direct connection of the chip onto the strip detector is under investigation starting with the bump bonding technology and joining the worldwide effort on 3D vertical interconnection.